\documentclass{optica-article}

\journal{opticajournal} 

\articletype{Research Article}

\usepackage{graphicx}
\usepackage{subcaption}
\usepackage{lineno}

\begin{document}

\title{Classification of Single Photons in Higher-Order Spatial Modes via Convolutional Neural Networks}

\author{Manon P. Bart\authormark{1,2}, Sita Dawanse\authormark{1}, Nicholas J. Savino\authormark{1}, Viet Tran\authormark{3,4}, Tianhong  Wang\authormark{3,4}, Sanjaya Lohani\authormark{5}, Farris Nefissi\authormark{3}, Pascal Bass\`ene\authormark{3,4}, Moussa N'Gom\authormark{3,4,**}, and Ryan T. Glasser\authormark{1,*}}

\address{\authormark{1} Department of Physics and Engineering Physics, Tulane University, New Orleans, Louisiana 70118, USA
\authormark{2} NASA Goddard Space Flight Center, Greenbelt, Maryland 20771, USA

\authormark{3} Department of Physics, Applied Physics, and Astronomy, Rensselaer Polytechnic Institute, 110 Eighth St, Troy, New York 12180 USA

\authormark{4} Center for Biotechnology and Interdisciplinary Studies, Rensselaer Polytechnic Institute, 1623 15th St, Troy, New York 12180, USA

\authormark{5} Department of Electrical and Computer Engineering, Southern Methodist University, Dallas, Texas 75205, USA}

\email{\authormark{*}rglasser@tulane.edu} 
\email{\authormark{**}ngomm@rpi.edu} 

\begin{abstract*} 
Spatial modes are a promising candidate for encoding information for classical and quantum optical communication due to their potential high information capacity. Unfortunately, compensation of the wavefront upon propagation through the atmosphere is necessary to benefit from advantages spatial modes offer. In this work, we leverage the success of convolutional networks in denoising and classifying images to improve information transfer of spatial modes. Hermite-Gauss, Laguerre-Gauss, and Ince-Gauss modes are experimentally generated using single photons and imaged. A denoising autoencoder corrects for turbulence effects on the wavefront, followed by a convolutional neural network to classify mode orders. The model achieves a 99.2 $\%$ classification accuracy across all modes, and Hermite-Gauss modes exhibited the highest individual mode accuracy. As the convolutional networks rely solely on intensity, they offer an efficient and cost-effective tool for optical communication systems in the single photon limit.
\end{abstract*}

\section{Introduction}
There are many technical advantages of optical communication through free space, such as large bandwidths, narrow beam divergence, and smaller weight and power requirements than radio frequency links \cite{FSO} . While several degrees of freedom are possible to send information through optical channels, the spatial mode of light is promising due to higher information capacity and larger bit-transfer per photon, \cite{BenefitsOAM2,BenefitsOAMChanCap1}, mutual orthogonality \cite{BenefitsOAM1,BenefitsOAM3} and increases in security at both the classical and single photon level \cite{BenefitsOAM1}.  If spatial modes are manifest in the form of higher order Gaussian modes, for example: Hermite-Gaussian (HG) modes, Laguerre-Gaussian (LG) modes, and helical Ince-Gauss (IG) \cite{Bandres:04, Bandres:04a, pampaloni2004gaussian}, information can be contained in the order numbers. In this case, the ability to distinguish order numbers is paramount. As with any optical beam propagating through free space links, this becomes difficult due to diffraction from long range propagation, turbulence in the atmosphere, and other noise which ultimately decreases the transmission probability of optical signals  \cite{FSO, FSO1,FSO2, OptSat_Noise0,turb1,AO1}. 

Due to this, communication channels with spatial modes are an experimental challenge that require further analysis on atmospheric effects on the wavefront as well compensation of wavefront distortion. While the statistics of turbulence are well studied \cite{SC2,SC3,turb3}, the random fluctuations in intensity lead to difficulty in predicting and mitigating the effects of turbulence on the signal after propagation. Machine learning has been proposed as a way to minimize effects of turbulence in optical channels \cite{App_OptTurb0,App_TurbCorr1,App_TurbCorr2,App_TurbCorr3} and optical communication in free space \cite{App_OptCom0,App_OptCom1,App_OptCom2,App_OptCom3,App_OptCom_gan0} and fiber \cite{App_OptCom_fib0}. Convolutional neural networks (CNNs) and convolutional autoencoders (CAEs) are particularly useful due to their ability to extract features and patterns in large sets of images  \cite{ml_patterns}. Several methods have been proposed using CAEs \cite{App_CAEComm1} and CNNs to aid in improvement of optical communication schemes \cite{App_CNNComm2,App_CNNComm3} and technology \cite{App_CNNTech1}. Given this, we expect them to be a viable method for classifying higher-order spatial modes.  

In our experiment, three different spatial modes, HG, LG and helical IG, are generated through computer generated holography and passed through simulated levels of turbulence which is implemented using spatial light modulators (SLM). The spatial modes are imaged and used as input to the convolutional network. The convolutional network is trained to classify the mode order of the image, where the final network pipeline consists of a denoising CAE followed by a classifying CNN. We first train a CNN which classifies the mode and order of the experimentally generated images. Then, as autoencoders have proven effective in improving final accuracy when partial image corruption is present \cite{caeMotivation}, a denoising CAE is trained to generate turbulence-corrected images that will ultimately be used prior to the CNN. While work has been done in the application of machine learning in spatial mode recognition \cite{App_OptCom0}, we examine the classification accuracy at the single photon limit.

\section{Experiment}
\begin{figure}[h!]
    \centering
    \includegraphics[width=.9\linewidth]{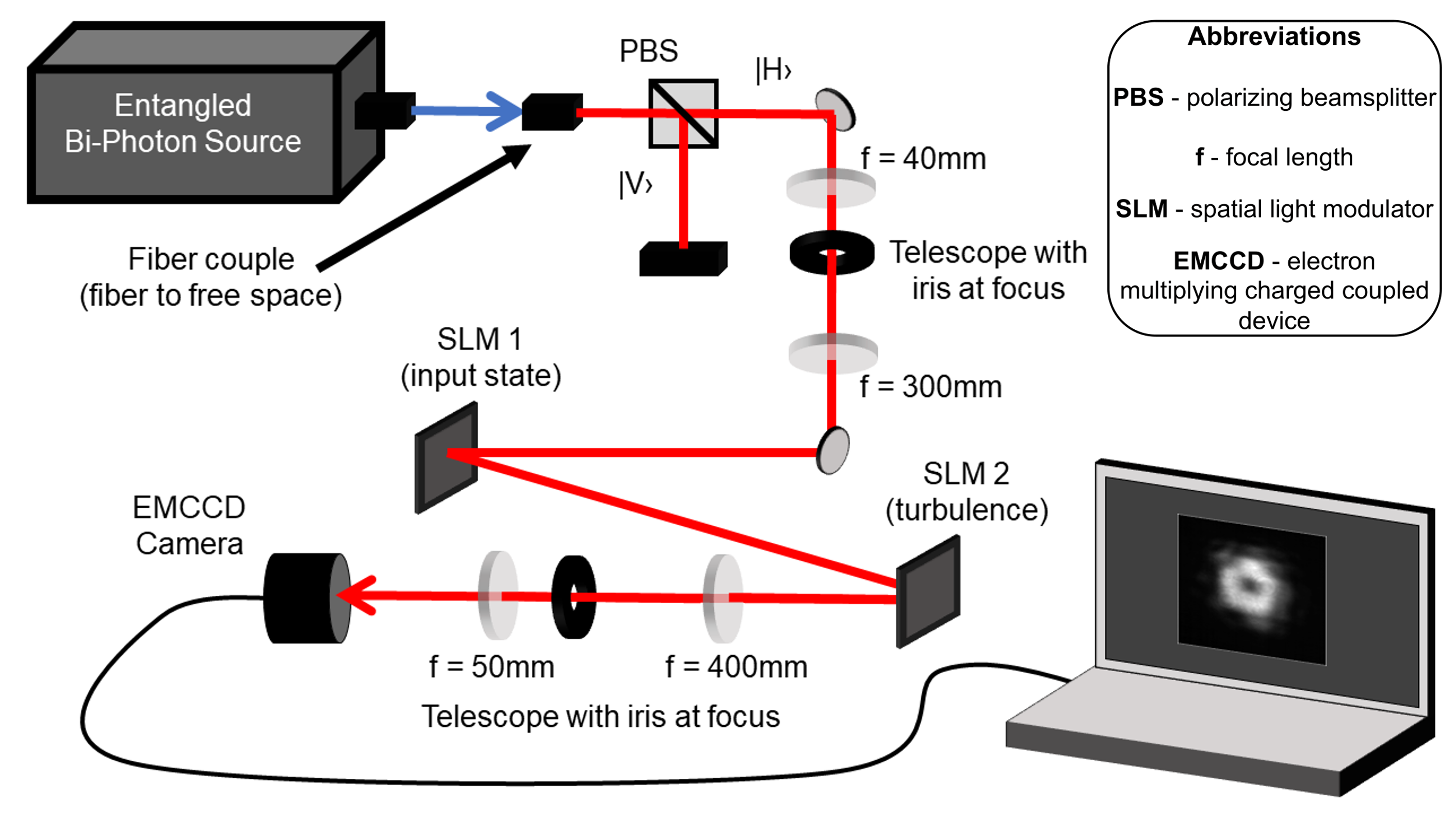}
    \caption{Experimental setup used to generate higher order spatial modes with single photons, pass the modes through turbulence, then measure them with a single photon resolving detector.}
    \label{fig:single_photon_turb_mode_setup}
\end{figure}

We experimentally generated the propagation of higher order spatial modes of light. A diagram of the setup used is shown in Figure \ref{fig:single_photon_turb_mode_setup}. Entangled photon pairs are emitted from a Qubitekk Entangled Photon Source with Integrated Laser Bi-Photon Source via a fiber optic cable. The photons are generated via parametric down conversion and exit the source in polarization entangled pairs. The wavelength of the source is 810nm. We select the horizontally polarized photon, from each pair using a PBS. The beam of single photons is then passed through a telescope with a magnification of 7.5. An iris as the focal point of the telescope is used to trim any spatial abnormalities and ensure a gaussian beam is incident.

The spatial modes are generated using two Holoeye Pluto-2 SLMs. The first SLM is programmed to transform the incident Gaussian beam into a higher order spatial mode. The second SLM is programmed to model turbulence through the atmosphere. The turbulence is modeled by a Kolmogorov phase with Von Karman spectrum effects \cite{Kolmo_PM, App_TurbCorr2}. 

\begin{equation}
    \Phi(\kappa) = 0.023\text{r}_{0}^{-5/3}(\kappa^2+\kappa_0^2)^{-11/6} e^{-\kappa^2/\kappa_{\text{m}}^2}
\end{equation}

The Fried parameter is $\text{r}_{0}=(0.423\text{k}^2\text{C}_{\text{n}}^2\text{z})^{-3/5}$. The wavevector is defined by k=2$\pi/\lambda$, z is the propagation distance, and $\kappa$ is the spatial frequency, $\kappa_{\text{m}} = 5.92/l_0$, and  $\kappa_0=2\pi/L_0$ where $l_0$ and $L_0$ are the inner and outer scales of turbulence, respectively. C$_{\text{n}}^2$ quantifies the strength of the atmospheric turbulence. The generated spatial mode orders, as well as the discrete turbulence levels, are listed in Table \ref{tab:Parameters for Turbulence Simulation and Imaging}. Several experimental factors led to a higher C$_{\text{n}}^2$ than typically observed; however, machine learning's performance is inherently robust to the specific statistical properties of the noise, thus the analysis on denoising and classification benefits remain valid for the range of image corruption chosen. Further information can be found in the Supplemental Document. LG$_{n,l}$ modes are described by the radial number, \textit{n}, and the azimuthal number, \textit{l}. HG$_{n_x,n_y}$ are described by $n_x,n_y$, which correspond to the number of nodes in cartesian coordinates of the spatial profile, and the helical IG$_{p,m}$ = IG$_{p,m}^{even}$ - \textit{i}IG$_{p,m}^{odd}$ is described by order \textit{p} and degree \textit{m}.

\begin{table}[htbp]
    \centering
    \caption{\bf Parameters for Experimental Setup}
    \begin{tabular}{|l|l|}
        \hline
        \textbf{LG mode orders} & LG$_{4,0}$,LG$_{1,1}$,LG$_{4,2}$,LG$_{0,4}$,LG$_{2,4}$  \\
        \hline        
        \textbf{HG mode orders} & HG$_{0,4}$,HG$_{1,2}$,HG$_{2,6}$,HG$_{4,4}$,HG$_{4,6}$  \\
        \hline  
        \textbf{IG mode orders} & IG$_{3,1}$,IG$_{4,4}$,IG$_{8,4}$,IG$_{9,1}$,IG$_{10,2}$  \\
        \hline
        \textbf{C$_{\text{n}}^2$ (x10$^{-4}$ m$^{-2/3}$)} & 30, 10, 7, 3, 1  \\
        \hline
    \end{tabular}
    \label{tab:Parameters for Turbulence Simulation and Imaging}
\end{table}

Since a beam of single photons is examined, an exposure time much greater than the inverse photon emission rate is used in order to resolve the photons spatial profile. We varied the exposure time for the EMCCD, from 0.01 to 1 second. For a single mode and order, 1,250 1024x1024 pixel images are generated in total. As the spatial modes were concentrated in the center of the EMCCD, the images were cropped to 200x200 pixels to decrease the computational time of the neural network. After collection of the images, a CNN is trained to classify the order of the different spatial modes. 

\section{Machine Learning Methods}
The full architecture consists of a CAE to denoise images, followed by a CNN to classify mode orders. The model is trained on half of the experimental data, with the other half reserved for prediction. Further details on training parameters and results for both networks are contained in the supplemental document.

\subsection{Convolutional Neural Network}
\begin{figure}[ht!]
    \centering
    \includegraphics[width=.85\linewidth]{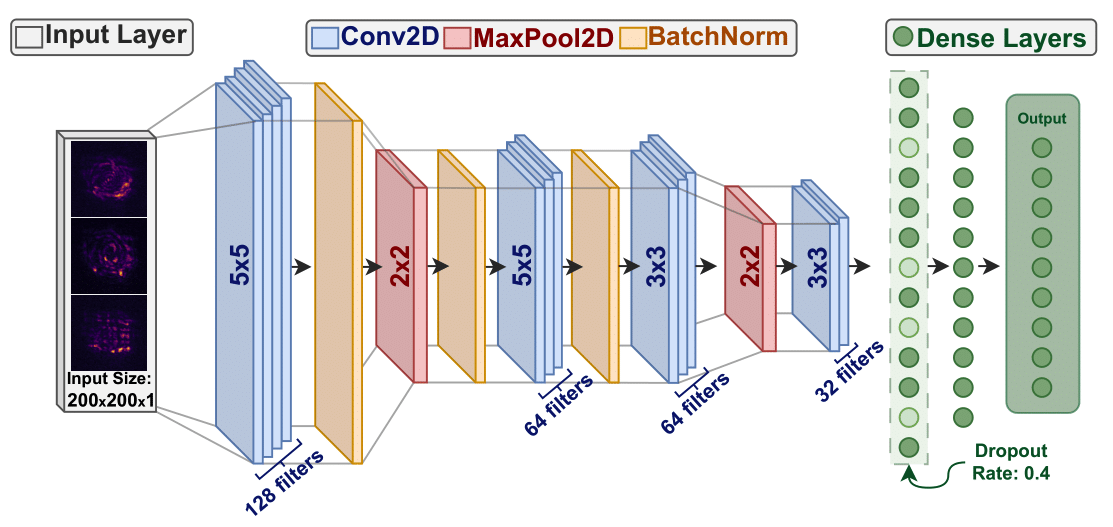}
    \caption{Model architecture for the convolutional neural network (CNN). The type of layer in the architecture is defined at the top of the figure. The kernel matrix size is listed inside the convolutional and max pooling layers. The input layer takes in images of pixel size 200x200 and the output layer corresponds to the different spatial mode orders.}
    \label{fig:CNNArchitecture}
\end{figure}

The CNN is trained to classify the mode and order of the receiver images. The supervised CNN takes an input of 200x200 pixel images, followed by convolutional layers to learn features of the images, and dense layers to classify the mode orders at the output. Our CNN architecture is shown in Figure \ref{fig:CNNArchitecture}. 

The model consists of an input convolutional layer using 128 filters with kernel size of 5x5, employing ReLU activation and a stride of 2. Following this, two batch normalization layers are inserted with a max pooling layers of kernel size 2x2 in between the two. There are three more convolutional layers all with ReLU activation, as well as several batch normalizations and max pooling layers of the same kernel size. The pooling layers reduce the dimensionality \cite{cnn}. Batch normalization is included to accelerate and stabilize training \cite{batchnorm}. After the convolutional steps, the output is vectorized to be input into dense layers. There are two fully connected dense layers with 128 and 64 neurons, respectively, both employing ReLU activation, and a dropout layer of rate 0.4 is used in between to minimize overfitting \cite{dropout}. The final dense layer uses softmax activation to generate class probabilities. The model classified the modes and orders with 97.702 $\pm$ .005\%  accuracy with a loss from the sparse categorical cross entropy loss function of 0.074 $\pm$ 0.014. The uncertainty corresponds to the first standard deviation of the accuracy and loss.

\subsection{Convolutional autoencoder for denoising images}
\begin{figure}[htbp!]
    \centering
    \includegraphics[width=.85\linewidth]{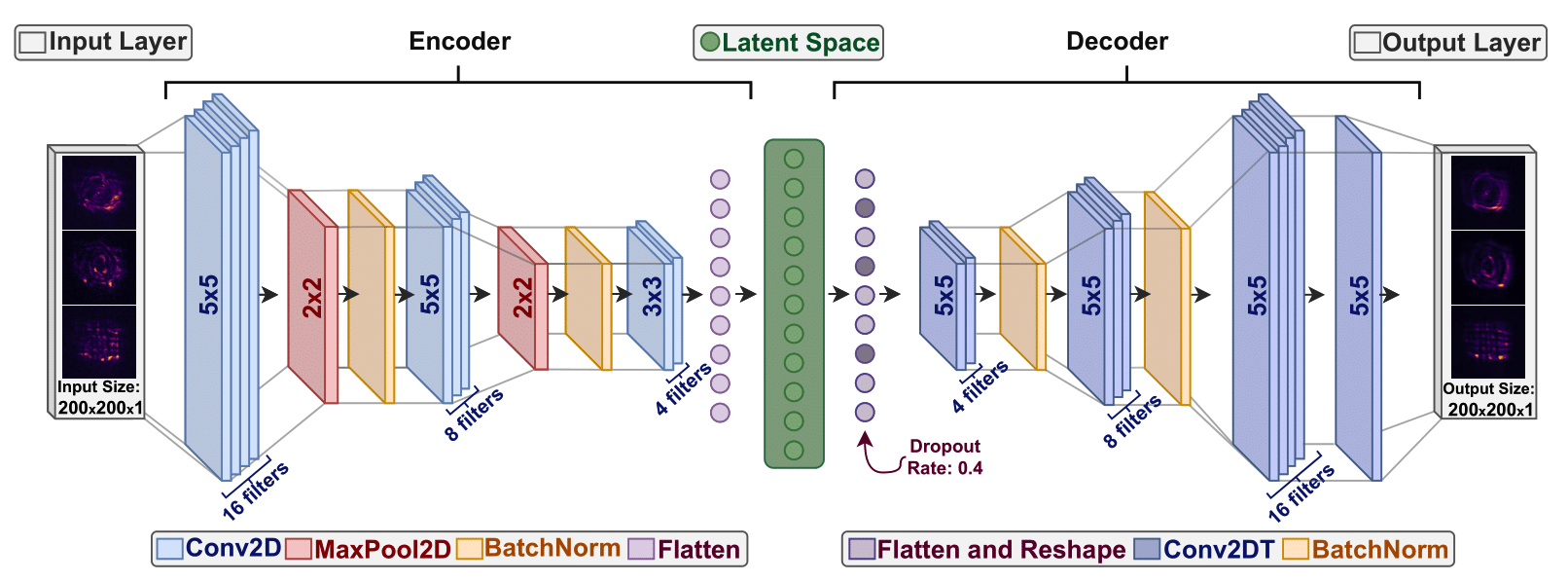}
    \caption{The model architecture for the convolutional autoencoder (CAE). The encoder and decoder layers are shown in the bottom left and right legend, respectively. The latent space is of vector size 30. The matrix inset in the convolution and max pooling layers correspond to the kernel size.}
    \label{fig:CAEArchitecture}
\end{figure}

Denoising CAEs are unsupervised models that transform the input into an output of same dimension with little distortion \cite{cnnApp1}. The CAE maps the experimental images to a denoised representation, resulting in a corrected image with minimized effects of turbulence. The CAE architecture is shown in Figure \ref{fig:CAEArchitecture}.

\begin{figure}[ht!]
    \centering    \includegraphics[width=.8\linewidth]{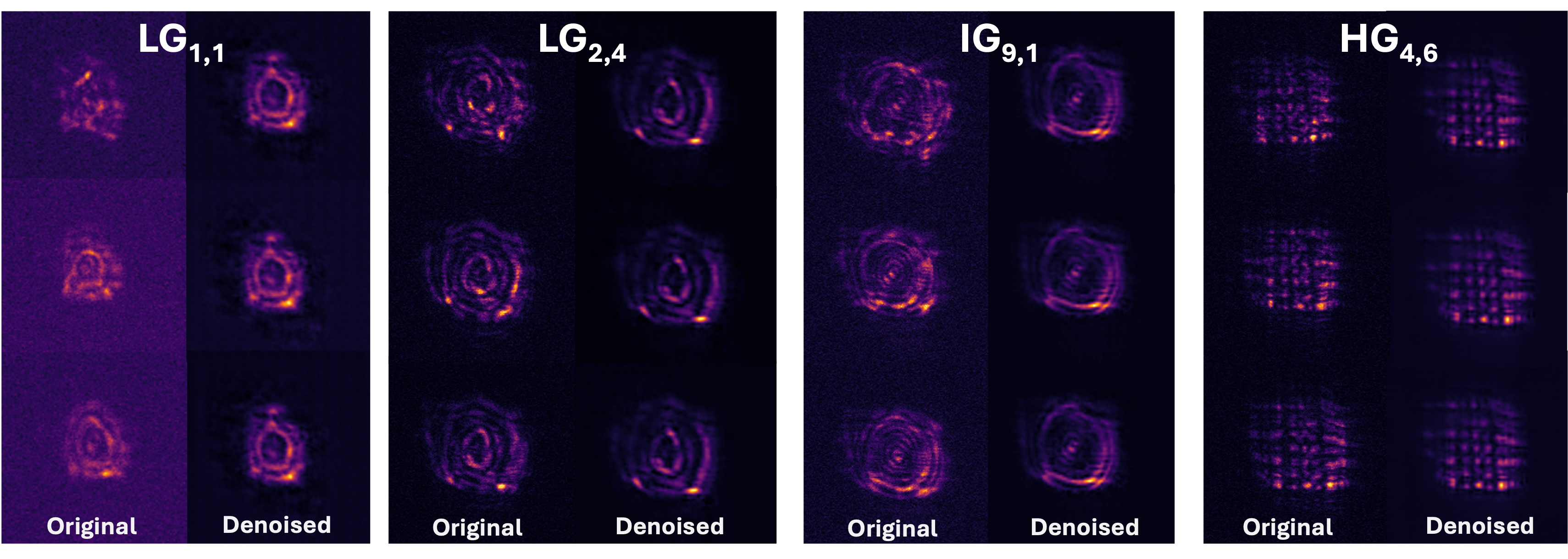}
   \caption{Original experimentally generated images and corresponding denoised images at the output of the CAE for C$_{\text{n}}^2$ (x10$^{-4}$ m$^{-2/3}$) = 30, 10, 7 from top to bottom.}
    \label{fig:samplecaeresults}
\end{figure}

A CAE consists of an encoder, latent space, and decoder. The encoder uses convolutional layers to extract main features from the data. The encoder consists of three convolutional layers with 16, 8 and 4 feature mappings of kernel size 5x5 and stride length of 2. The first two convolutional layers are followed by a maxpooling layer of kernel size 2x2. The output of the encoder is flattened leading to a vector, latent space, of size 30. The latent space vector retains the most important features. During decoding, transposed convolutional layers expand the latent space representation, reconstructing the input data while preserving the learned features \cite{caeHowto}. The decoder has an inverted architecture as the encoder. All convolutional layers use a rectified linear unit (ReLU) activation function. The decoder activation, at the output of the autoencoder, is linear. A mean-squared-error (MSE) loss function is used for training. A near order of magnitude reduction in the MSE was demonstrated for all spatial modes. Various experimental images and their representation with reduced turbulence effects are shown in Figure \ref{fig:samplecaeresults}. 

Slight deviations of the experimental images from the theoretical profile are attributed to generation of light by phase-only hologram without separating diffractive order and the tilted angle of the SLM. As all mode orders are effected by this, it is not corrected for by the CAE. This is in contrast to the random spatial variations due by turbulence, which are corrected by the CAE. Further details are explained in the supplemental material. 

\section{Results}
\begin{figure}[htbp!]
    \centering
    \includegraphics[width=\linewidth]{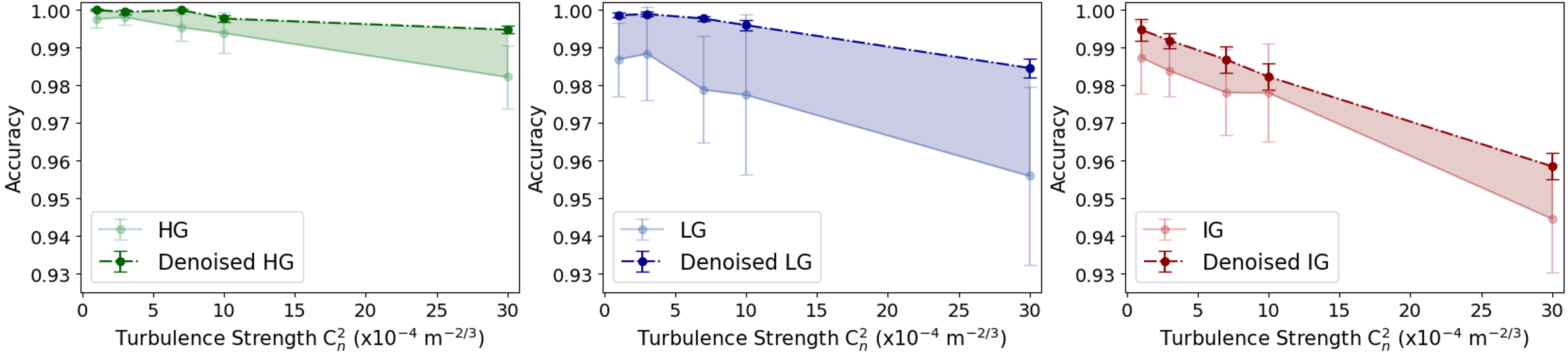}
    \caption{Average classification accuracy for each spatial mode for a range of turbulence levels with (dotted line) and without (solid) a convolutional autoencoder (CAE). The shaded region represents the improvement from utilizing the CAE.  Error bars constitute one standard deviation from the mean for 10 trials. }
    \label{fig:singlephotontestacc}
\end{figure}

After training is complete, the unseen experimentally generated images were tested for classification accuracy with and without the inclusion of the CAE. The resulting classification accuracy and sparse categorical cross entropy loss for utilizing solely the CNN is 98.2 $\pm$ 0.9 $\%$ and 1.57 $\pm$ 0.90, which corresponds to the mean and one standard deviation of the mean across 10 trials. Inclusion of the CAE prior to the CNN increased the classification accuracy to 99.209 $\pm$ 0.1 $\%$. The loss was reduced to 0.21 $\pm$ 0.17. In addition to testing the final accuracy across all modes and turbulence, the performance of the model with and without the CAE are compared for individual spatial modes and turbulence level in Figure \ref{fig:singlephotontestacc}.  

\begin{figure}[ht!]
    \centering
    \includegraphics[width=\linewidth]{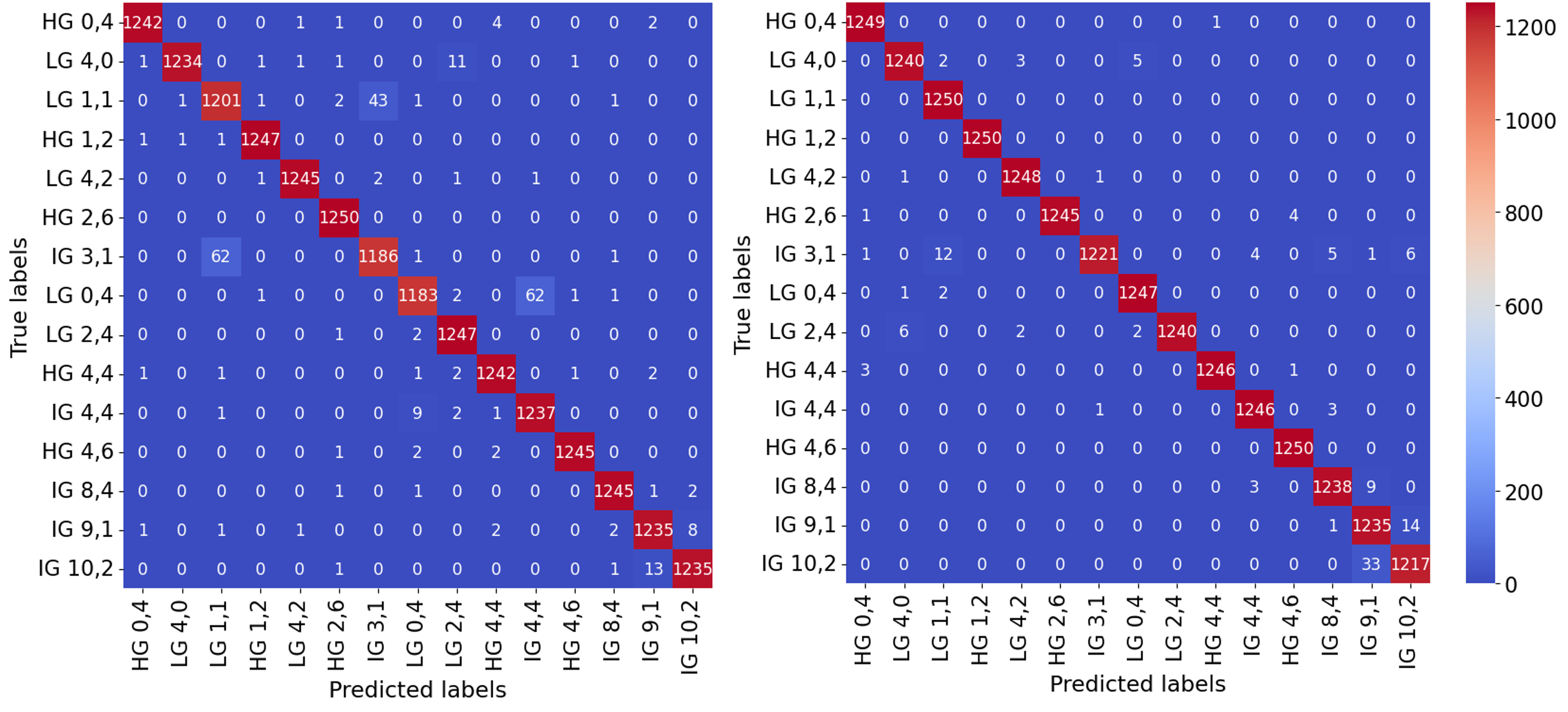}
    \caption{Confusion matrix for all spatial modes and orders without the use of a convolutional autoencoder (CAE) --left and with the use of a CAE --right.}
    \label{fig:singlephotontestcm}
\end{figure}

The CNN with and without the CAE demonstrated excellent performance in classifying spatial modes for all levels of turbulence. As expected, accuracy decreased as we increased the turbulence strength. We found the HG modes outperformed the LG and IG modes for all ranges of turbulence. The inclusion of a CAE prior to inputting the images into the classifying CNN improved the accuracy for all spatial modes, largely in the case of low turbulence. The largest improvement in accuracy was demonstrated by LG modes. The inclusion of the CAE also significantly reduces the variance of accuracy of the convolutional network over multiple trials. Figure \ref{fig:singlephotontestcm} contains a confusion matrix to visualize the resulting classification errors across all turbulence strengths with and without the CAE.

The most common errors occurred between LG and IG modes. Potential causes of errors can be attributed to physical relations between LG modes and IG modes. The ellipticity parameter is a dimensionless quantity which is defined by $\epsilon = 2f_0^2/w_0^2$ , where the parameters $f_0$ and $w_0$ scale the physical size of the mode. IG modes transition to LG modes as $f_0$ approaches 0. In this case, the indices of both modes are related by LG$_{n,l}$ = IG$_{p=2n+l,l=m}$. This is reflected in our experimental results, where $\epsilon = 2$, by the classification errors between IG$_{31}$ and LG$_{11}$, as well as IG$_{4,4}$ and LG$_{0,4}$. The inclusion of the CAE prior to the CNN mitigates the errors in these cases. Common errors with inclusion of the CAE are present between different higher order IG modes. HG modes with and without the aid of the autoencoder are robust to imaging classification for all ranges of turbulence, which suggests the spatial structure of the image may lead to a cleaner wavefront after propagation through turbulence. 

\section{Conclusion}
The primary objective of this work was to evaluate the effectiveness of machine learning in spatial mode order classification. The convolutional network architecture, which consists of a CAE followed by a CNN, achieved a classification accuracy of 99.2 $\pm$ 0.1 $\%$ across all modes and orders. HG modes exhibited the highest classification accuracy, suggesting the spatial features of HG are better preserved after turbulence propagation. The denoising CAE identified distinguishing features of each mode order, and generated a turbulence-robust image representation, which was notably useful when strong turbulence impacts the classification accuracy due to decoherence. In addition to the increase in classification accuracy, the CAE decreased the variance in classification accuracy over multiple trials, leading to a consistent and reproducible tool for classification of the various modes. Since the model is pretrained, it is integrable with other tools used to demodulate signals, such as adaptive optics, while not significantly increasing the system's size, weight, and power. In addition, our model solely relies on intensity measurements from the camera, which makes the convolutional networks straightforward and reproducible in training and implementation. As utilizing spatial modes as information carriers to perform classical and quantum optical communication relies on a clean wavefront and turbulence introduces random intensity fluctuations during propagation, the results in this work suggest convolutional networks support information transmission in the single photon limit.

\section{Back matter}
\begin{backmatter}
\bmsection{Funding} National Science Foundation Graduate Research Fellowship (2139911); P.B., and M.N’G. acknowledge the support by a grant 442 from the Gordon and Betty Moore Foundation to the PAIR-UP Imaging Science Program. R. T. G. and M.N’G. acknowledge the support of the U.S. Department of Energy (DOE) under Grant No DE-SC0024676.

\bmsection{Disclosures} The authors declare no conflicts of interest.

\bmsection{Data Availability Statement} Data underlying the results presented in this paper are not publicly available at this time but may be obtained from the authors upon reasonable request.

\bmsection{Supplemental Document}
See Supplement 1 for supporting content.

\end{backmatter}

\bibliography{sample}

\end{document}


\maketitle

\section{Training and validation of the convolutional networks}
Training and validation of the convolutional neural network (CNN) and convolutional autoencoder (CAE) was implemented on half of the experimentally generated images, which corresponds to 9,375 200x200 pixel images. The other half of the experimentally generated images were used for testing our models.

\subsection{Classifying Convolutional Neural Network}
\begin{figure}[htbp!]
    \centering
        \includegraphics[width=.8\textwidth]{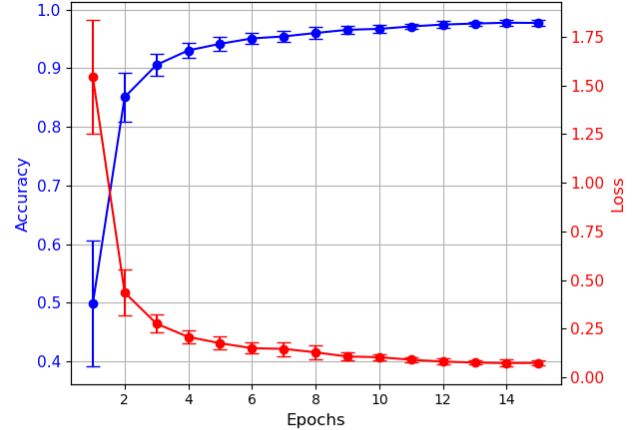}
        \caption{Average training accuracy and loss for classifying all higher order spatial modes. The model was run for 15 epochs and repeated for 10 trials to ensure the repeatability of the model. The final training accuracy is 97.702 $\pm$ .005\%  accuracy. The final sparse categorical cross entropy loss is 0.074 $\pm$ 0.014.}
        \label{fig:cnntrainingaccloss}
\end{figure}

The CNN used an 80/20$\%$ split for training and validation. The model was compiled using the Adam optimizer with a learning rate of 0.001, and the sparse categorical cross-entropy function is used as the loss function for 15 epochs. A grid search was utilized to determine the optimal initializers, regularizers and regularizer strength for every layer. The regularizers tested in our grid search include [L1,L2,None] with a regularization strength varied from 0.1, 0.01, and 0.001. The initializers tested include [GlorotUniform, HeNormal, None]. Our final hyperparameters are GlorotUniform as the kernel initializer and no kernel regularization was used. The training loss and accuracy are shown in Figure \ref{fig:cnntrainingaccloss}.

We conducted training for 10 trials, in order to ensure the optimized model architecture was repeatable. The model classified the modes and orders with 97.702 $\pm$ .005\%  accuracy with a loss from the sparse categorical cross entropy loss function of 0.074 $\pm$ 0.014. The uncertainty corresponds to the first standard deviation of the accuracy and loss.

\subsection{Denoising Convolutional Autoencoder}
\begin{figure}[htbp!]
    \centering
    \includegraphics[width=\linewidth]{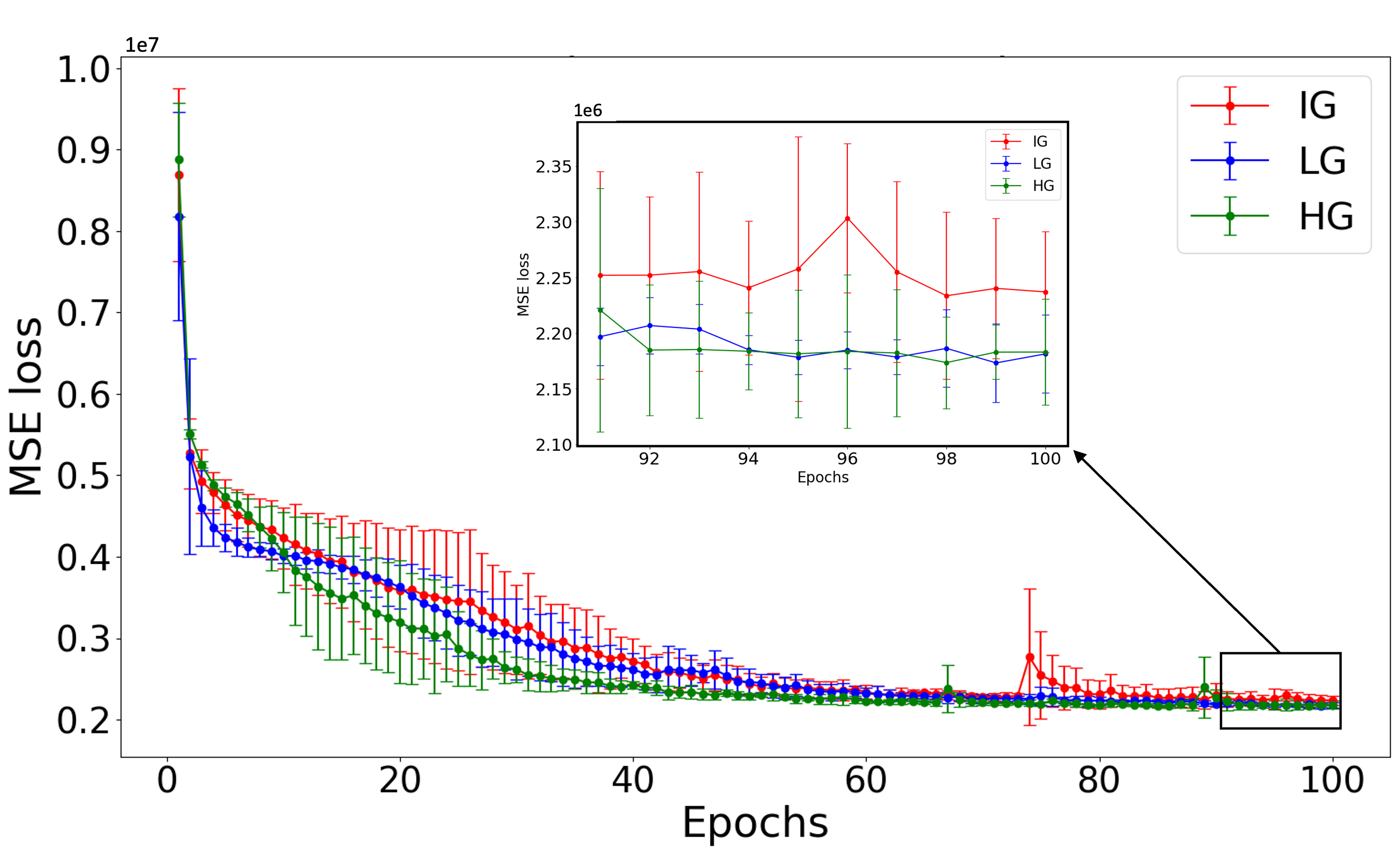}
   \caption{Training loss for the convolutional autoencoder (CAE) for each spatial mode. The Adam optimizer was used for training and mean-squared-error (MSE) was used as the metric for loss. Training was conducted for 100 epochs for 5 trials. The error bars constitute one standard deviation of the mean loss over the trials. (inset) The last ten epochs of training.}
    \label{fig:caetrainresults}
\end{figure}

The optimized CAE architecture was trained on each spatial mode separately, with 70$\%$ of the images in the training set and 30$\%$ in the validation set. The loss function used was mean-squared-error (MSE) and is trained with the Adam optimizer with a learning rate of 0.2. The CAE was trained for 100 epochs, and the resulting loss versus epochs for the training is shown in Figure \ref{fig:caetrainresults}. 

The results were averaged over 5 trials, where the number of epochs was chosen by when the MSE converged and there was little variance in the final MSE. 

\section{Generation of Experimental Images}
In order to generate and image spatial modes passing through turbulence using computer generated holography, the propagation of single photons needs to cover the active area of the spatial light modulator (SLM). To ensure this, single photons were replaced with coherent laser light and the propagation was analyzed. The Gaussian beam begins at approximately 2 mm in diameter, and after passing through a telescope, the beam is slightly greater than 1.5 cm in diameter. The beam size is increased such that it covers the entire active area of the SLM, which is approximately 1 cm across. The beam is then focused by our last lens into the camera to fully form. A diagram of the full experimental set up is shown in Figure 1. 

\begin{figure}[htbp!]
    \centering
    \includegraphics[width=.9\linewidth]{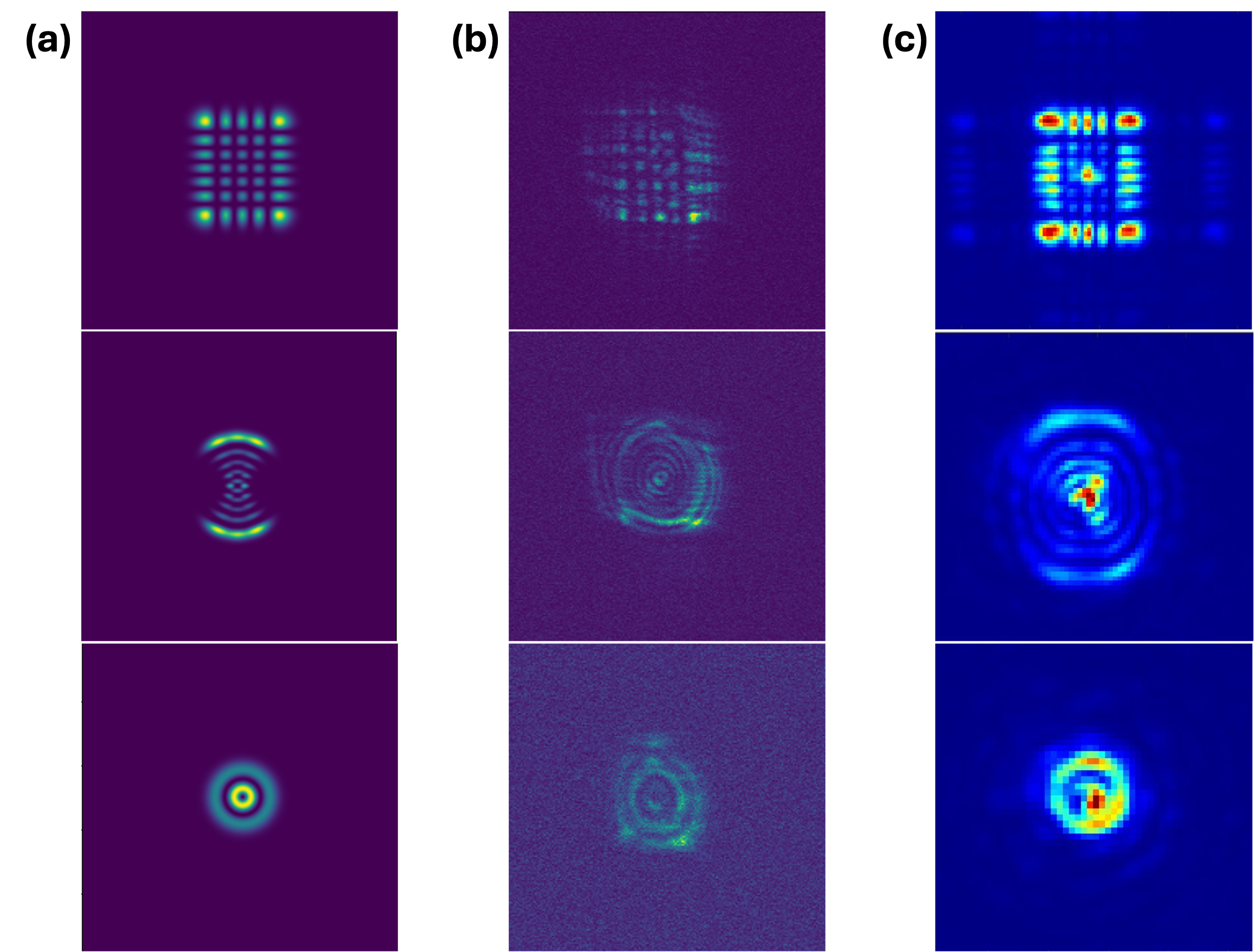}
   \caption{Experimentally generated images of HG$_{4,6}$, helical IG$_{9,1}$, and LG$_{1,1}$, from top to bottom. (a) Simulated propagation of spatial modes without turbulence. (b) Experimentally generated images used for training and validation of the CNN and CAE at lowest turbulence strength, C$_{\text{n}}^2$ (x10$^{-4}$ m$^{-2/3}$) = 1. (c) Experimentally generated images without turbulence.}
    \label{fig:theorysim}
\end{figure}

While this process ensures the coverage of the active area on the SLM for both the spatial modes and turbulence generation, this process also leaves unmodulated light in the beam. Due to space restrictions which inhibit us to separate the diffractive orders, the structured light was generated using a phase-only mask in the zeroth order. This factor, along with a slight tilted angle in the SLM, leads to small differences in the theoretical profile and experimentally generated images. Comparison of our experimentally generated images and the theoretical profile are shown in Figure \ref{fig:theorysim}.

Figure \ref{fig:theorysim} (b) depicts the experimentally generated images used for our convolutional networks at the lowest turbulence strength, C$_{\text{n}}^2$ (x10$^{-4}$ m$^{-2/3}$) = 1. In comparison with the theoretical spatial mode profile without any turbulence, as shown in Figure \ref{fig:theorysim} (a), brighter intensities in the outer edge of the beam profile, elongated edges, rings and the separation of topological charges can be observed. These effects can be attributed to the lack of separation of the diffractive orders and the tilted angle of the SLM. This is shown to a lesser degree in experimental images generated without turbulence, as shown in Figure \ref{fig:theorysim} [c]. As every spatial mode experiences these astigmatism effects, the accuracy in classification for a machine learning model is not affected.

Finally, due to the small size of the spatial modes and large distances, z = 25 m, used to model the Kolmogorov phase with Von Karman spectrum effects, a large $C^2_n$ value was needed to generate the partial corruption of the images. The beam size, which was slightly greater than 1.5 cm, followed by propagation through the active area of the SLM, approximately 1 cm across, led to generated spatial modes which represent less than 0.2 cm of the field. As we are dealing with a large beam size in which the spatial modes cover a small area, the spatial modes are less affected by the turbulence and thus the need for compensation by a larger $C_n^2$ value \cite{Turb1,Turb2}. Although observed atmospheric turbulence is typically of smaller order, the main goal of the experiment was to generate noisy images. Further, the denoising CAE and classifying CNN focus on pattern recognition between classes of images and are impartial to the statistics or types of noise generated. The discrete $C_n^2$ values used in this work thus achieved generation of a range of images at different noise levels, through which classification for various levels of image corruption could be compared. 


\bibliography{supplement}